\documentclass[reprint, amsmath, amssymb, aps]{revtex4-1}
\usepackage{amsmath,amssymb,graphicx,mathrsfs,amsfonts}

\usepackage{newfloat}
\DeclareFloatingEnvironment[fileext=lop]{photo}

\newcounter{lastnote}

\newcommand{\ket}[1]{\ensuremath{\left|  #1 \right\rangle}}

\begin{document} 

\title{Superradiance on the milliHertz linewidth strontium clock transition}

\author{Matthew A. Norcia}
\affiliation{JILA, NIST, and University of Colorado, 440 UCB, Boulder, CO  80309, USA}
\author{Matthew N. Winchester}
\affiliation{JILA, NIST, and University of Colorado, 440 UCB, Boulder, CO  80309, USA}
\author{Julia R. K. Cline}
\affiliation{JILA, NIST, and University of Colorado, 440 UCB, Boulder, CO  80309, USA}
\author{James K. Thompson}
\affiliation{JILA, NIST, and University of Colorado, 440 UCB, Boulder, CO  80309, USA}

\date{\today}

\begin{abstract}

Today's best atomic clocks are limited by frequency noise on the lasers used to interrogate the atoms.  A proposed solution to this problem is to create a superradiant laser using an optical clock transition as its gain medium.  This laser would act as an active atomic clock, and would be highly immune to the fluctuations in reference cavity length that limit today's best lasers.  Here, we demonstrate and characterize superradiant emission from the mHz linewidth clock transition in an ensemble of laser-cooled $^{87}$Sr atoms trapped within a high-finesse optical cavity.  
We measure a collective enhancement of the emission rate into the cavity mode by a factor of more than 10,000 compared to independently radiating atoms.  We also demonstrate a method for seeding superradiant emission and observe interference between two independent transitions lasing simultaneously.  We use this interference to characterize the relative spectral properties of the two lasing sub-ensembles.  
\end{abstract}

\maketitle

\subsection*{Introduction:}


Optical atomic clocks have recently achieved fractional instability in their ticking rate of a few parts in $10^{18}$ \cite{Katori2015, Bloom2014, Hinkley2013, chou2010frequency}.  Significant improvements in the accuracy, precision and bandwidth of clocks and the lasers used to probe them would significantly advance a broad range of science and technology including: tests of general relativity \cite{chou2010optical}, proposed gravitational wave detection \cite{2015arXiv150100996L}, searches for variations of fundamental constants\cite{PhysRevLett.98.070801} and new gravitational couplings \cite{PhysRevLett.100.140801}, searches for dark matter\cite{arvanitaki2015searching,NaturePhys933}, gravitational potential sensing for geodysy \cite{CHR10}, stabilization of future quantum networks\cite{QNetworkClocks2014}, and explorations of quantum many-body physics\cite{PhysRevLett.113.120402,Zhang19092014}.  


At the heart of these optical clocks are atoms like $^{87}$Sr, which has a quantum state with a long decay lifetime of roughly 150 seconds \cite{porsev2004hyperfine,santra2004properties}.
The inverse lifetime of this state corresponds to a frequency linewidth of 1~millihertz, which is  more than $10^9$ times narrower than typical optically excited states.  
This linewidth relative to the 
frequency of the optical photon emitted when the atom decays corresponds to a large fundamental quality factor $Q=4\times 10^{17}$, which is a key figure of merit for a clock. 
However, because of frequency instability in the lasers used to probe the atoms, today's best clocks can only resolve a much broader linewidth, and therefore a lower $Q$ --- the atoms are more precise than the lasers used to measure them \cite{Hinkley2013}.

For decades, heroic efforts have been made to reduce the frequency linewidth of conventional lasers by stabilizing their frequency to mechanically stable optical reference cavities  \cite{YCI99,kessler2012sub}.  The primary limitation of this approach is thermal Brownian motion of the cavity mirror spacing that produces noise in the cavity's resonance frequency \cite{numata2004thermal,notcutt2006contribution}. 
Here we present the first key step toward a radically different approach to narrow linewidth lasers --- directly collecting light emitted from a long-lived quantum state \cite{CHE09,MYC09}.  
Such a laser would be of order $10^4$ times less sensitive to thermal and technical sources of cavity frequency noise \cite{bohnet2012steady,norcia2015cold}.  

In this approach, the long lifetime becomes a serious challenge.  Typically, photons are emitted far too slowly to serve as a useful phase or frequency reference and are emitted into all directions, making them difficult to utilize.  To overcome these limitations, we achieve pulsed superradiant lasing for the first time on an ultra-weak optical clock transition:  the mHz linewidth $^3$P$_0$ to $^1$S$_0$ clock transition at 698~nm in  $^{87}$Sr.  Superradiant stimulation of photon emission allows us to efficiently collect photons emitted from the 150~second lifetime state in under 100~ms. The emitted laser light both serves as an absolute frequency reference and offers a new path towards lasers with linewidths at or below the mHz level \cite{MYC09}, orders of magnitude narrower than has previously been achieved with traditional optical reference cavities \cite{kessler2012sub}.  


\begin{figure*}[tp]
\includegraphics[width = 6.75in]{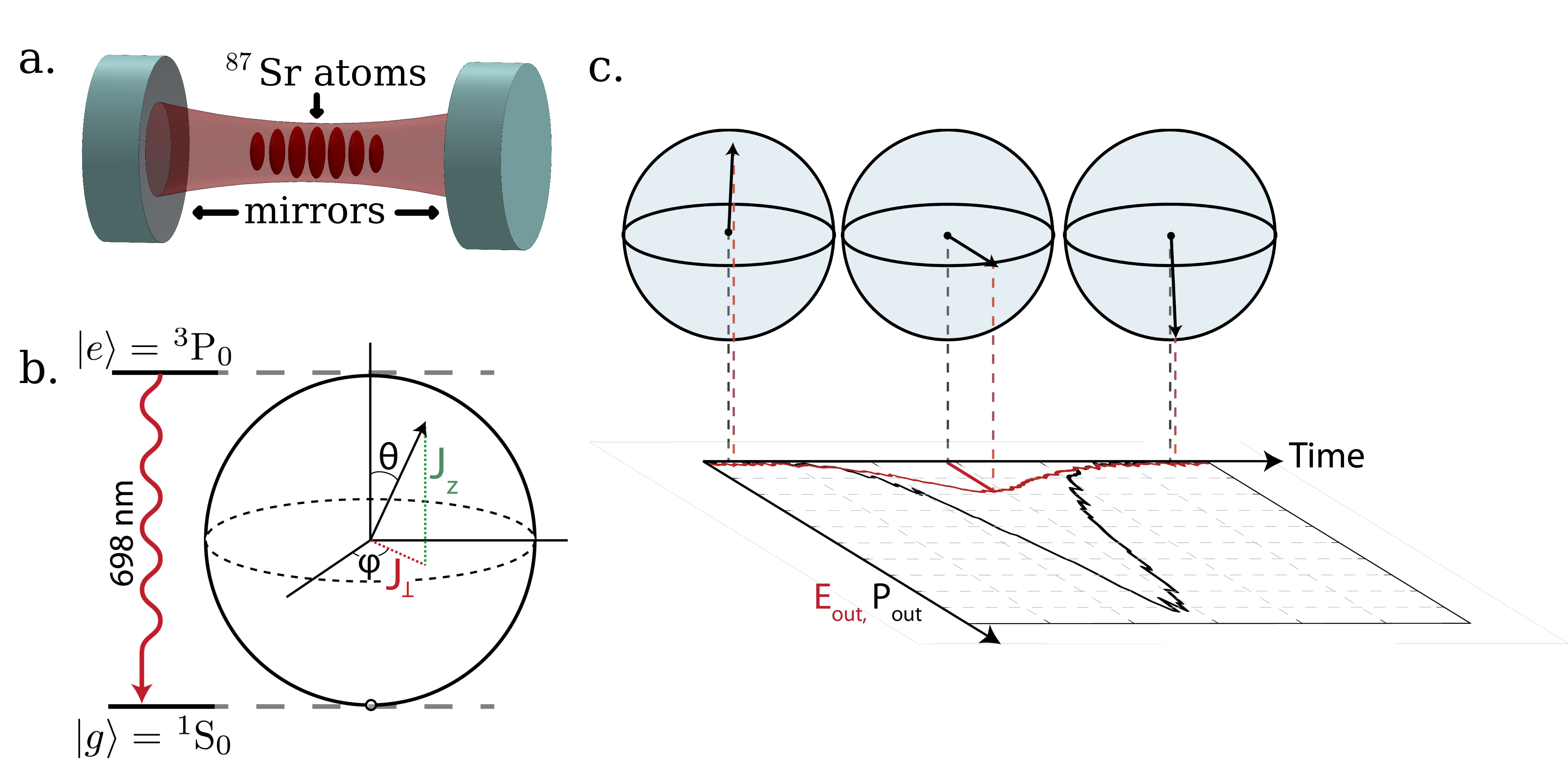}
\caption{Experimental overview  (a) Our system consists of an ensemble of up to $N = 2.5 \times 10^5$ laser-cooled $^{87}$Sr atoms confined in a magic wavelength optical lattice within a high finesse ($F = 2.4 \times 10^4$, linewidth $\kappa = 2 \pi \times 160$~kHz) optical cavity.  (b) The atoms undergo pulsed superradiant lasing on the $^3$P$_0$ to $^1$S$_0$ optical clock transition that has a natural decay time of 150~seconds, or an equivalent linewidth of 1~millihertz. The state of the atomic ensemble can be represented by a collective Bloch vector, explained in the text.  (c)  A representation of a superradiant pulse on the Bloch sphere.  The Bloch vector behaves like a highly damped pendulum that starts inverted at the north pole of the Bloch sphere (excited state).  Quantum fluctuations disturb the system from its unstable equilibrium position, causing the Bloch vector to swing down the Bloch sphere, emitting peak radiation at the equator and ultimately relaxing to the south pole (ground state) as inversion is lost.  The radiated electric field (red trace) is proportional to the perpendicular projection of the Bloch vector, $J_\perp$, which at its peak is proportional to $N$.  The radiated power (black trace) is proportional to the square of the radiated electric field, and at its peak is therefore proportional to $N^2$.  This is one way to understand the origin of the collective enhancement in emission rate. The black output power trace on the projection plane is actual data. 
}
\label{fig:ExpDiag}
\end{figure*}

In order to achieve lasing, the collectively enhanced emission rate from the atoms must be made larger than atomic decoherence rates, a stringent requirement for this ultra-weak transition.    To increase the collectively enhanced decay rate, we trap the atoms within a high finesse optical cavity (Figure 1a), effectively increasing the optical depth of the atomic ensemble. To suppress atomic decoherence, we rely on techniques used to provide long coherence times in optical lattice clocks \cite{PhysRevLett.91.173005, Ye1734} --- by laser cooling and confining the atoms along the cavity axis with a magic-wavelength optical lattice, we eliminate first-order Doppler shifts in the direction of emission without imposing large shifts to the lasing transition frequency.  

In conceptually related work, Raman transitions between ground hyperfine states of rubidium have enabled proof of principle explorations of lasing in the deep bad-cavity or superradiant regime\cite{bohnet2012steady, BCWHybrid}, in which the bandwidth of the laser's gain medium (the atomic transition) is much narrower than that of the laser's optical cavity.  However, because the frequency stability of a Raman laser is limited by the stability of the lasers used to induce optical decay between ground states, such a system is not suitable for a frequency reference.  For this, a true narrow-linewidth optical transition is required.  
In addition, the clock transition used here is orders of magnitude narrower than the effective decay linewidth used in Raman systems.  



Superradiance has been studied in a variety of other more broadband systems, including thermal molecular and atomic gasses \cite{skribanowitz1973observation, gross1976observation}, Rydberg atoms \cite{gross1979maser}, atoms trapped near photonic crystals \cite{goban2015superradiance}, ions \cite{PhysRevLett.114.023602}, artificial atoms \cite{scheibner2007superradiance}, and other Raman systems \cite{thompson2006high, PhysRevLett.92.213601, chaneliere2005storage}.  
More generally, collective interactions mediated by the 7.5~kHz linewidth dipole-forbidden $^3$P$_1$ to $^1$S$_0$ transition in $^{88}$Sr have been studied in the context of lasing, cavity QED, and collective scattering \cite{norcia2015strong, norcia2015cold, PhysRevLett.114.093002, kwong2014cooperative, bromley2016collective}.  
This work pushes into a new regime, exploring collective interactions mediated by a transition nearly seven orders of magnitude weaker than even that weak transition.


\subsection*{Experimental System}

Our experimental system consists of up to $N = 2.5 \ \times \ 10^5$ $^{87}$Sr atoms cooled to $10\ \mu$K and tightly trapped along the axis of a high-finesse ($F = 2.4 \times 10^4$, linewidth $\kappa = 2\pi \times 160$~kHz) optical cavity by an optical lattice.  The lattice is near the magic wavelength of 813.4274~nm, for which the frequency shift of the two clock states is equal, making the transition frequency independent of lattice intensity \cite{PhysRevLett.91.173005}.  The interaction between the atoms and cavity mode can be characterized by the cooperativity parameter $C$ of cavity QED.  In our system, the peak single-particle cooperativity parameter is $C = 0.41$ (assuming a Clebsch Gordan coefficient of 1).  This number represents the relative probability that an atomic excitation leaves the system as a photon transmitted through a cavity mirror versus into free space.  For a collective excitation of $N$ atoms, this ratio is enhanced to $\sim NC$.  For our typical atom numbers, this means that an atom is far more likely to emit a photon into the cavity mode than into free space.

The state of the atomic ensemble can be represented by a collective Bloch vector, which is the vector sum of the Bloch vectors of the $N$ individual atoms. This is illustrated in Figure 1b.  
An atom in the excited state, $\ket{e}$ ($^3$P$_0$), has a Bloch vector pointing up, while an atom in the ground state, $\ket{g}$ ($^1$S$_0$), has a Bloch vector pointing down.  
An atom in an equal superposition of $\ket{e}$ and $\ket{g}$ corresponds to a Bloch vector on the equator of its Bloch sphere, with phase $\varphi$ determined by the phase of its superposition $\ket{g}+e^{i \varphi}\ket{e}$. 
To account for the spatial distribution of the atoms, the phase of each atom is defined relative to the phase of a cavity mode resonant with the $\ket{e}$ to $\ket{g}$ transition, evaluated at the location of each atom.  
A collective Bloch vector on the equator of the Bloch sphere corresponds to each atom in a superposition of $\ket{e}$ and $\ket{g}$ with the appropriate phases to collectively radiate into the cavity mode.  

The atoms radiate an electric field into the cavity at a rate proportional to $J_\perp$, the magnitude of the projection of the Bloch vector onto the equatorial plane of the Bloch sphere.  
The collective enhancement of emission results from the fact that the power radiated is proportional to the square of the electric field.  Because the electric field is proportional to the atom number $N$, the radiated power scales with $N^2$ \cite{Gross1982}.  
The electric field radiated into the cavity acts on the atoms by causing rotation of the Bloch vector about an axis in the equatorial plane of the Bloch sphere by a rate proportional to $\sqrt{M_c}$, where $M_c$ is the average number of photons in the cavity.  This is the mechanism of stimulated emission in the superradiant regime.  
For an ensemble whose Bloch vector lies in the northern hemisphere of the Bloch sphere, this leads to positive feedback for emission.  The atoms will radiate into the cavity, and the radiated electric field will then cause the Bloch vector to tip further from the north pole, and thus to radiate more strongly.  The result is a pulse of light that builds up gradually, reaches a peak in power as the Bloch vector passes the equator, and falls to zero as the atoms reach the ground state (Figure 1c).

\begin{figure}[!htb]
\includegraphics[width=3.5in]{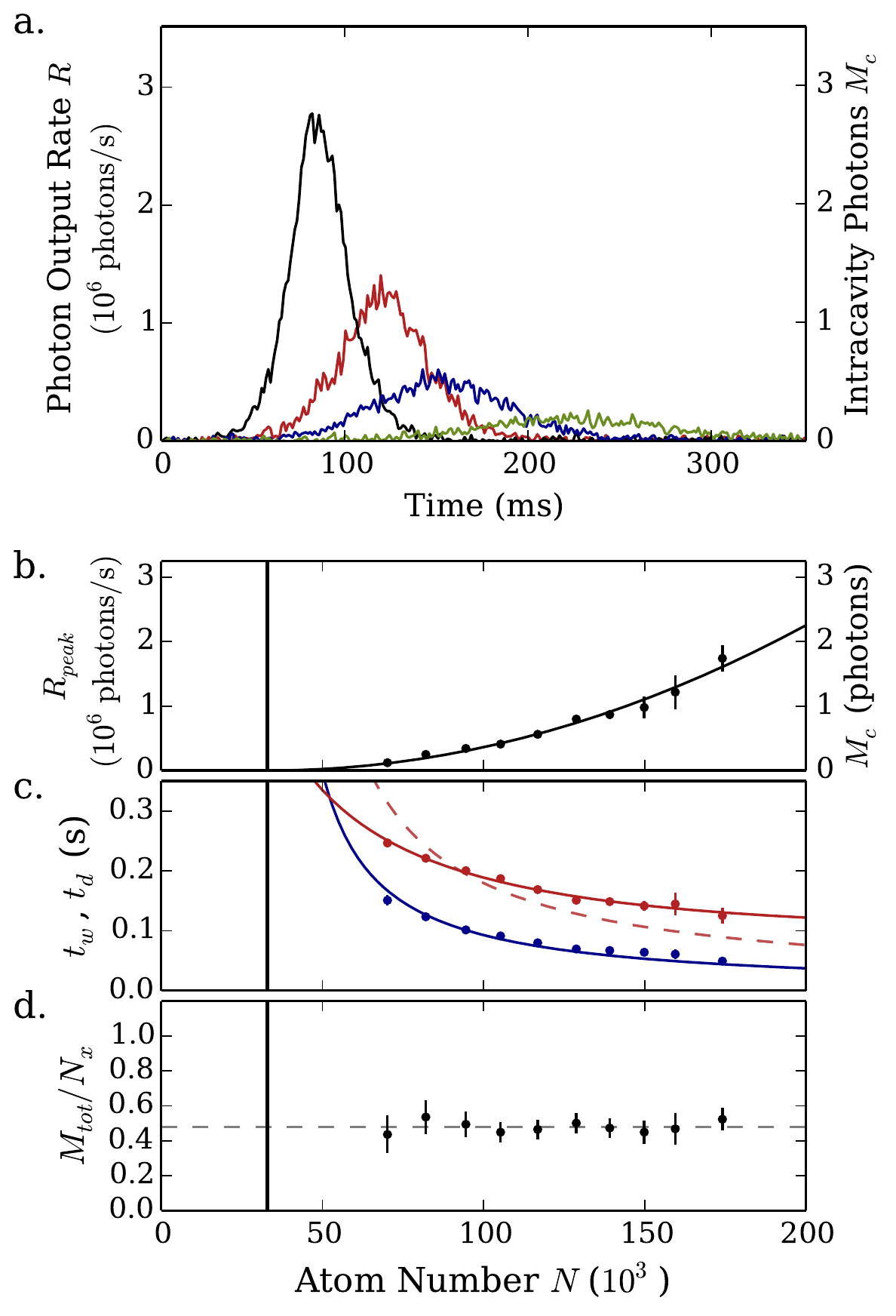}
\caption{Spontaneously generated superradiant pulses.  (a)  Representative single time traces of photon output rate $R$ for pulses at different atom number $N\approx$ $ 100\times10^3$ (green), $ 125 \times 10^3$ (blue), $150 \times 10^3$ (red), $ 200 \times 10^3$ (black).  The equivalent average intracavity photon number is calculated on the right as $M_c= R/\kappa$.
(b) Peak photon output rate, $R_{peak}$, versus initial total atom number.  The black line is a quadratic fit to the output power.  We observe a horizontal offset in the fit, indicating a threshold atom number $N_t$ (black vertical line in b,c,d).  The threshold results from decoherence and atom loss and is a signature of lasing that would not be present for single-atom emission.  (c)  FWHM pulse duration (blue) and delay of peak power (red) versus initial total atom number.  The blue line is a fit to the predicted functional form for the pulse duration, with $N_t$ determined from the fit to  $R_{peak}$.  The dashed red line is a fit to the pulse delay, assuming threshold is due to atomic homogeneous broadening without atom loss.  The solid red line is a fit to the pulse delay, assuming the threshold is set only by atom loss from the lattice.  (d)  The ratio of emitted photons $M_{tot}$ to the number of atoms in excess of threshold $N_x$, plotted versus atom number. The dashed line is the average ratio, showing that 48\% of the atoms in excess of threshold  participate in the superradiant pulse, largely independent of $N$.
}
\label{fig:ExpDiag}
\end{figure}

\subsection*{Observation of Superradiant Pulses}
To observe superradiant pulses, we prepare atoms in $\ket{e}$, the nuclear  $\ket{F=9/2, m_f = 9/2}$ sub-level of $^3$P$_0$.  We first optically pump the atoms to $\ket{g}$, the  $\ket{F=9/2, m_f = 9/2}$ sub-level of $^1$S$_0$, then adiabatically transfer up to 75\% of the atoms to $\ket{e}$ using a frequency-swept 698~nm transfer beam applied through the cavity.  To prepare the atoms with full inversion (no atoms in $\ket{g}$) and to ensure that the laser pulses are initiated by quantum noise rather than residual atomic coherence associated with the adiabatic transfer process, we then briefly apply lasers to the dipole-allowed $^1$S$_0$ to $^1$P$_1$ transition to heat any atoms remaining in the ground state out of the lattice. The state-preparation process is described in more detail in the supplemental material.


When all atoms are initially prepared in $\ket{e}$, we observe collectively enhanced decay on the clock transition.  
Both quantitative and qualitative features of the collectively enhanced emission are dramatically different from that of independent atoms.  
Not only does the collective enhancement lead to an emission rate into the cavity mode of up to $10^4$ times greater than that of independently emitting atoms, but the functional form of the decay versus time is distinctly non-exponential.  Figure 2a shows the photon output rate $R$ for four representative pulses recorded with different initial atom numbers.  Because the rate of collectively enhanced emission per atom scales with $N$, for higher atom numbers the pulses appear sooner, have shorter duration, and have a higher peak power than for lower atom numbers.

Figure 2b shows the characteristic $N^2$ scaling of the peak output power $R_{peak}$ versus atom number that one expects for superradiance.  In the presence of decoherence or atom loss, the atom number must exceed a threshold $N_t$ for superradiance to occur. $N_t$ is set by the requirement that the collectively enhanced decay rate exceeds the atomic decoherence rate.  Above this threshold, we predict $R_{peak} = \frac{1}{4 \xi} N_x^2 C \gamma$, where $N_x= N-N_t$ is the total number of atoms in the lattice $N$ in excess of the theshold atom number.  The inhomogeneous coupling of the atoms to the cavity mode is accounted for by the numerical factor $\xi \approx 2.95$. (see supplemental material for details and note that all following expressions account for this inhomogeneity.) From a fit of this form, we extract a fitted threshold of $N_t = 3.3 \times 10^4$ atoms.  From the fitted $N_t$ and known $C$ and $\gamma$, the measured peak photon emission rate $R_{peak}$ is 0.7(4) times the above predicted rate.

The time duration of the superradiant pulse provides a measure of the collectively enhanced decay rate.  The measured FWHM versus atom number $N$ is shown in Figure 2c (blue points).  We predict that the FWHM duration $t_w$ of the pulse is given by $t_w\approx 7.05 /(N_x C \gamma)$, such that the enhanced decay rate scales linearly with the excess atom number $N_x$.  We fit this functional form (blue line) to the data, with the threshold held fixed to $N_t = 3.3 \times 10^4$ atoms from above.  From this fit we find that the measured FWHM is 1.4(7) times the predicted FWHM given the known $C$ and $\gamma$.  

The measured delay time $t_d$ of the peak in output power versus atom number $N$ is shown in Figure 2c (red points).  In the presence of homogeneous broadening of the atomic transition, but with no atom loss, we expect the delay time to be given by $t_d \approx \frac{2(\ln{N} + \gamma_e)}{N_x C \gamma}$  (dashed red fit line with $N_t$ fixed and $C$ fitted), where  $\gamma_e \approx 0.577$ is the Euler-Mascheroni constant.  In the presence of atom loss from the lattice at rate $\gamma_\ell$, we observe a delay time in numerical simulations of the form $t_d \approx \alpha \gamma_\ell+\frac{2(\ln{N} + \gamma_e)}{N C \gamma}$ where $\alpha$ is a constant, a functional form which seems to better describe the data (solid red fit line with $N_t$ fixed and $C$ and $\alpha$ fit.)    

We define the number of atoms that participated in a superradiant pulse in terms of the integrated number of photons  $M_{tot}$ emitted from the cavity mode, i.e. one photon equals one participating atom.  Figure 2d shows the number of emitted photons per atom in excess of threshold, $M_{tot}/N_x$.  
We observe that above threshold, $M_{tot}/N_x=0.48(15)$. Inhomogeneous coupling to the cavity mode would predict $M_{tot}/N_x=0.7$ partially accounting for the observed participation. Atomic collisions, which lead to an atom-number dependent contribution to $N_t$ may account for the additional reduction of participation. We may contrast this level of participation to the case where atoms emit independently: if there were no stimulation, only 0.1\% of atoms would emit a photon into the cavity mode during our measurement time. 

\begin{figure}[!htb]
\includegraphics[width=3.5in]{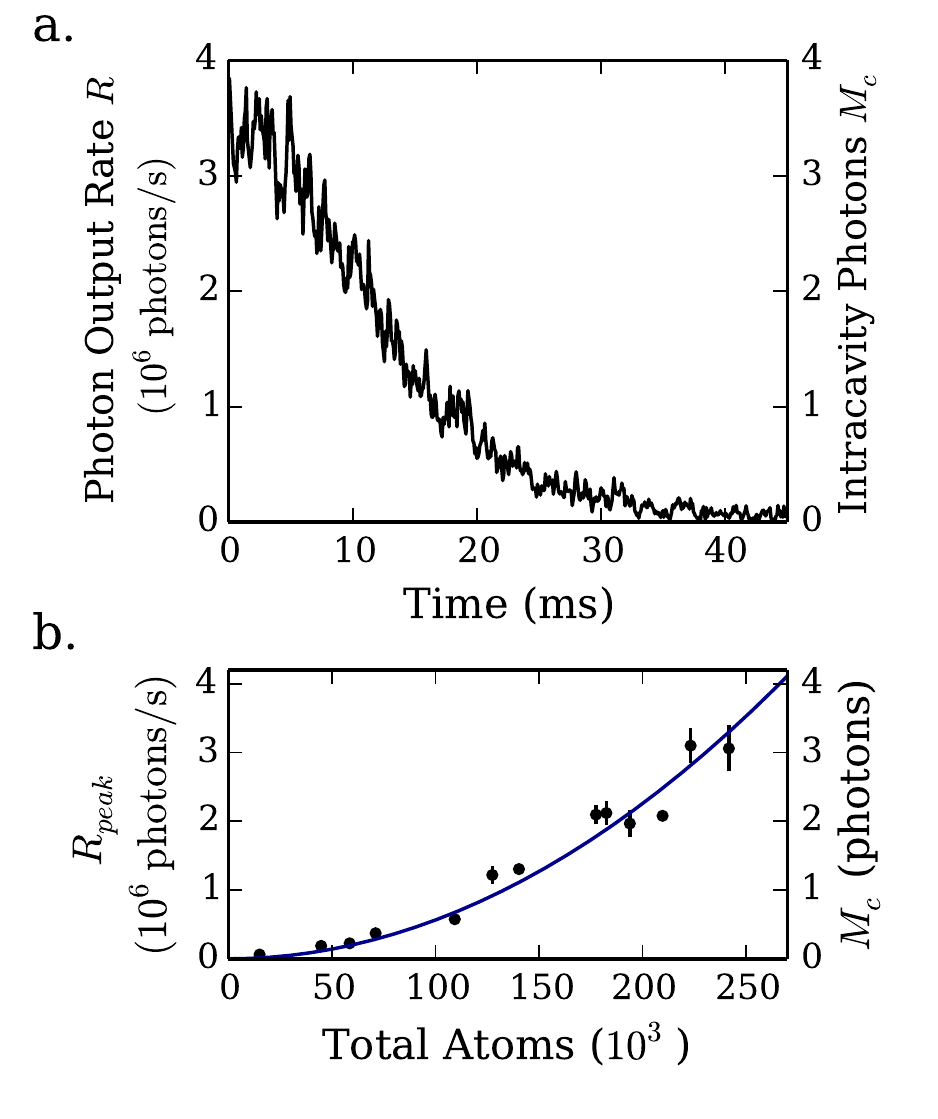}
\caption{Seeded superradiant pulses. (a) By terminating adiabatic transfer with atoms in a superposition of $\ket{e}$ and $\ket{g}$, we seed coherence in the atomic ensemble.  Here, the atomic Bloch vector is rotated to just above the equator (i.e. a small amount of initial inversion.)  Seeding leads to the immediate onset of superradiant emission, in contrast to the non-seeded pulses shown in Figure 2 for which quantum noise seeds the coherence.  (b) Peak output power for seeded pulses exhibits $N^2$ scaling.  In contrast to results of Figure 2, seeded pulses exhibit a peak photon output rate consistent with no threshold: $N_t=0$.}
\label{fig:ExpDiag}
\end{figure}

\subsection*{Seeding Atomic Coherence}

\begin{figure*}[!htb]
\includegraphics[width=7in]{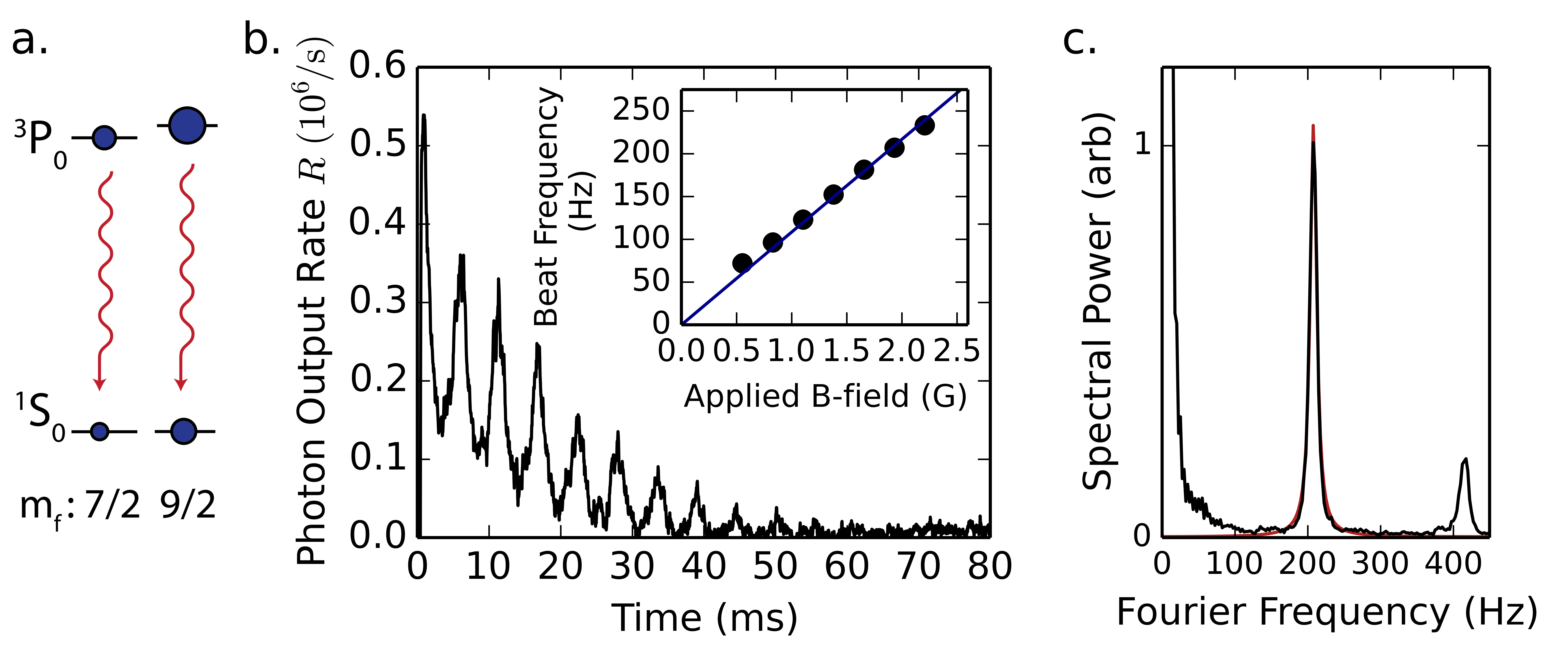}
\caption{Lasing on multiple transitions at the same time leads to beating in output power  (a) Atoms prepared in the $m_f = 9/2$ and $m_f = 7/2$ Zeeman sub-levels radiate simultaneously into the cavity.  (b) Interference between the electric fields radiated from the two transitions leads to a modulation of the output power.  An average of 20 time-traces is plotted, showing that the modulation has the same phase between trials of the experiment.  (c) We compute a Fourier transform of an averaged time trace, showing a peak at the frequency splitting of adjacent Zeeman sub-levels ($\approx 200$ Hz).  The peak at $\approx 400$ Hz indicates that a smaller number of atoms have been prepared in $m_f = 5/2$.  A Lorentzian fit to the average power spectrum (red line) returns a FWHM linewidth of 11 Hz, due to the finite length of the pulse.  The inset of (b) shows the center frequency of the Lorentzian fit versus applied magnetic field.  The blue line shows the predicted splitting between adjacent Zeeman sub-levels.  
}
\label{fig:fig4}
\end{figure*}

Instead of preparing the atomic ensemble in the excited state with no initial coherence as before, we can terminate the adiabatic transfer process early and prepare the atoms in a superposition of $\ket{g}$ and $\ket{e}$.  This seeds the collectively enhanced emission and unlike in the non-seeded pulses of figure 2, we detect an immediate output of light from the cavity. Figure 3a shows a typical output trace resulting from terminating the adiabatic transfer with the Bloch vector just above the equator of the Bloch sphere.  

Seeding the coherence in this manner also eliminates the threshold behavior observed in the spontaneously generated pulses of Figure 2.  Figure 3b shows the peak output power $R_{peak}$ versus $N$ for pulses seeded with the Bloch vector just above the equator of the Bloch sphere.  This data is well described by a quadratic fit with no offset, i.e. $N_t=0$.  

We view this technique for establishing collectively enhanced emission with no threshold or delay time to be a key tool for the development of superradiant sources. In this work, being able to use these signatures of collectively enhanced emission to incrementally tune the system to meet threshold was essential.  More fundamentally, seeding allows collectively enhanced emission to be achieved in systems that are incapable of meeting threshold, but that may still be of metrological value.  
If superradiant pulses are used to stabilize the frequency of another laser, seeding could be used to reduce dead-time that contributes to Dick noise aliasing \cite{Dick1987}.

\subsection*{Simultaneous Lasing on Multiple Transitions}

For all preceding data, we applied optical pumping to populate primarily the $^1$S$_0$, $m_f = 9/2$ sub-level before adiabatic transfer, resulting in a single relevant lasing transition.  
We can deliberately reduce the efficiency of the optical pumping to populate both the $m_f = 9/2$ and $m_f = 7/2$ ground states, and then adiabatically transfer the atoms into superpositions of ground and excited states with different $m_f$ projections, represented in Figure 4a.  This creates two separate sub-ensembles of atoms that interact with the same cavity mode but have slightly different transition frequencies \cite{2015arXiv150306464W, PhysRevLett.113.154101}.  
We observe a modulation in the output power at the magnetic-field induced frequency difference between the two transitions.  This modulation is the result of interference between the fields radiated by the atoms lasing on the two transitions.  Because more atoms are prepared in the $m_f = 9/2$ sublevel, the total field radiated never goes through zero and the contrast of the modulation is not full.  

Figure 4b shows the average of 20 time traces recorded under these conditions, illustrating that the phase of the modulation is the same between trials, a result of seeding coherence into the two transitions.  To verify that the observed amplitude modulation is the result of beating between adjacent Zeeman transitions, we compute a Fourier transform of the emitted power (Figure 4c) and fit the peak in the power spectrum that corresponds to the output power modulation.  The frequency of this peak is plotted against our applied magnetic field in the inset of Figure 4b.  The slope and offset of beat frequency are consistent with the expected Zeeman splitting between the $m_f = 9/2$ and $m_f = 7/2$ transitions.  The smaller peak near 400~Hz indicates that a smaller number of atoms have been left in the $m_f = 5/2$ state.  

A Lorentzian fit to the peak in the average power spectrum returns a FWHM linewidth of 11~Hz, primarily reflecting the finite length of the pulse.  Because many photons are detected in a trial of the experiment, we can fit the center of the Lorentzian peak with deviation much smaller than its width.  


Treating this as a differential frequency measurement of two lasers, we compute a fractional Allan deviation of $2.6 \times 10^{-15}$ at $\approx 2$ seconds, the repetition rate of our experiment.  
Because many sources of frequency errors that are common-mode to the two lasing transitions are not captured by this measurement, this number does not indicate the ultimate performance of the system as a frequency reference.  It does, however, reflect a bound on its quantum-limited instability.  
To more fully characterize the spectral properties of the emitted light, including its sensitivity to effects that are common to the two transitions used here, it will be necessary to perform a comparison with an independent narrow linewidth laser \cite{kessler2012sub}, which is a subject for future work.  

\subsection*{Conclusion and Outlook}
We have demonstrated that an ultra-narrow optical transition can be made to lase in a pulsed manner, with each atom emitting up to a single photon.  In the future, it will be advantageous to operate in a continuous manner, with pump lasers applied to return the atoms to the excited state and a means of replenishing atoms lost to heating or collisions.  An important property of a continuous superradiant laser is that the linewidth of the emitted light is not limited by the collectively enhanced decay rate, as would be the case for single-atom decay \cite{bohnet2012steady,norcia2015cold}.  Rather, the fundamental limit to the linewidth of the laser is of order $C \gamma$, resulting from phase diffusion of the cavity field due to single-atom emission into the cavity mode \cite{MYC09}.  


This work demonstrates that dramatic effects can result from collective interactions with an optical field, even when mediated by an optical transition so weak that it takes roughly $150$ seconds to decay without stimulation.  
These interactions lead to stimulated emission in a regime where the cavity field is much shorter lived than the coherence of the atomic ensemble, and open new avenues for the improvement of optical clocks, ultrastable lasers, and other atomic sensors along with their many applications.  
\subsection*{Acknowledgements}

We gratefully acknowledge useful conversations with Murray Holland, Jun Ye, Ana Maria Rey and Thomas Perkins, as well as technical assistance from John Robinson and Karl Mayer.  All authors acknowledge financial support from DARPA QuASAR, ARO, NSF PFC, and NIST. This work is supported by the National Science Foundation under Grant Number 1125844. 

\bibliography{ThompsonLab}

\begin{thebibliography}{10}
\bibitem{AOS_ref_note}
 Given as technical information only, and does not represent an endorsement by
  NIST.
\end{thebibliography}

\bibliographystyle{Science}

\section*{Supplemental Material}

\subsection*{Atom Cooling and Trapping}

A new ensemble of atoms is loaded into the optical lattice roughly every 2 seconds. 
An atomic source from AOSense, Inc. with integrated oven, Zeeman slower and two 2-dimensional magneto-optical traps (MOTs) provides a collimated beam of atoms with speeds around 50 m/s to the main experimental chamber \cite{AOS_ref_note}.  Roughly $10^7$ atoms are then captured and accumulated in a 3-dimensional MOT, and further cooled to mK temperatures.  These slowing and cooling stages use the dipole-allowed $^1$S$_0$ to $^1$P$_1$ transition at 461~nm (see Figure S1).  

To cool the atoms to a temperature compatible with loading into the optical lattice, we then form a second-stage MOT using the 7.5-kHz linewidth, dipole-forbidden $^1$S$_0$ $F = 11/2$ to $^3$P$_1$ $F = 11/2$ transition at 689~nm.  We find that by applying a saw-tooth frequency modulation with peak-to-peak frequency deviation of 2~MHz to the MOT beams, we can robustly load up to $3 \times 10^5$ atoms into the lattice at a temperature of 10$\mu$K, as determined by turning off the lattice and measuring time-of-flight expansion of the atoms.  The atoms are spread over roughly 1 mm along the cavity axis, corresponding to around 2000 occupied lattice sites.  

At 698 (813)~nm, the 4~cm long cavity mode has a waist size of 74 (80)~$\mu$m.  At a typical lattice depth of 100~$\mu$K, the frequency of axial (radial) motion in the trap is 170~kHz (270~Hz), giving a Lamb-Dicke parameter $\eta = 0.16$ in the axial direction \cite{Janik85}.  For the $\ket{e}$ to $\ket{g}$ transition studied here $C = 0.33$, and the single photon Rabi frequency is $2g = 2 \times 2\pi \times 3.7$~Hz for a maximally coupled atom.

\begin{photo}[!htb]
\includegraphics[width=3in]{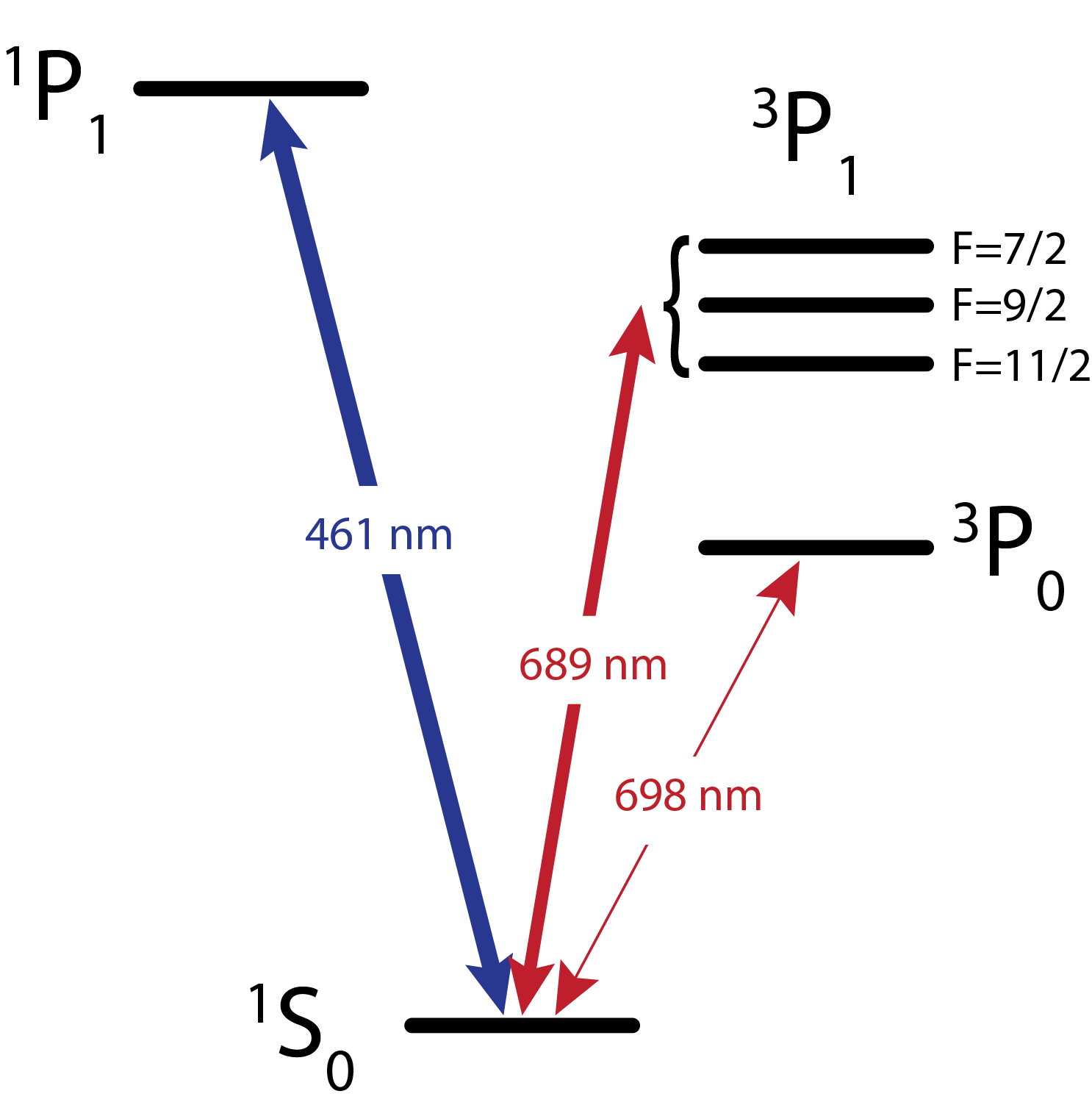}
\renewcommand\photoname{Figure}
\renewcommand{\thephoto}{S\arabic{photo}}
\caption{Energy level diagram of relevant transitions.  The dipole-allowed $^1$S$_0$ to $^1$P$_1$ transition (linewidth 32 MHz) is used for initial cooling and trapping.  The dipole-forbidden  $^1$S$_0$ to $^3$P$_1$, transition (linewidth 7.5 kHz) is used for final cooling into the optical lattice, and for optically pumping into the $^1$S$_0$ $m_f = 9/2$ Zeeman sublevel.  Superradiant lasing is observed on the dipole-forbidden $^1$S$_0$ to $^3$P$_0$, transition (linewidth 1~mHz) at 698~nm.  
}
\label{fig:fig4}
\end{photo}

\subsection*{State Preparation and Adiabatic Transfer}

Once atoms are loaded into the lattice, we apply a magnetic field of several Gauss perpendicular to the cavity to define the quantization axis.  We then typically optically pump atoms to the $m_f = 9/2$ state by applying a circularly polarized pump beam oriented along the magnetic field axis and near resonance with the $^1$S$_0$ to $^3$P$_1$, $F = 9/2$ transition.  

We adiabatically transfer the atoms to $^3$P$_0$ by applying a 698~nm laser sideband along the cavity axis.  The transfer sideband is coupled to a TEM00 mode of the cavity, and is linearly swept by 200~kHz in 20~ms over the mode of the cavity that is on resonance with the clock transition. 

The intensity of the 698~nm light at the location of individual atoms is spatially modulated by the standing wave formed within the cavity, with a peak Rabi frequency for atoms at antinodes of the cavity mode of roughly 5 kHz.  
We can transfer up to 75\% of the atoms to $^3$P$_0$.  Because the transfer beam is linearly polarized along the magnetic field, the atoms maintain their spin polarization $m_f$ when transferred to $^3$P$_0$. Because the transfer beam is applied parallel to the cavity axis, along which the atoms are tightly confined in the Lamb-Dicke regime, the interaction between the transfer beam and the atoms is Doppler-shift free.  The frequency sweep range is slightly greater than the cavity linewidth. Larger sweep ranges are found to perform less well, presumably because the frequency crosses an axial motional sideband transition frequency.

The adiabatic transfer technique has several advantages over other techniques of populating the excited state.  Compared to an incoherent pumping process where atoms reach $^3$P$_0$ from a higher-lying excited state, adiabatic transfer maintains the spin polarization present in the ground states and imparts far fewer photon recoils than would be required for optical pumping in $^3$P$_0$.  Compared to a resonant $\pi$ pulse, adiabatic transfer has the advantage of being less sensitive to the difference in Rabi frequencies experienced by different atoms and requires less stringent stabilization of the transfer laser frequency.  
Because the transfer beam is phase-matched to the cavity mode, terminating the transfer sweep near atomic resonance creates an atomic ensemble whose atoms are in superposition states with the correct phases to radiate into the cavity mode.  This allows us to prepare superradiant states, which emit collectively with no threshold or delay time.

\subsection*{Detection of Pulses and Atom Number}

To detect the superradiant laser pulses, the output of the cavity is coupled to a single-mode fiber and detected on a single-photon counting module (SPCM) whose TTL output is low-pass filtered to provide a signal proportional to the photon emission rate.  

After allowing a fixed time $T$ (typically 300~ms) to record the superradiant pulses, we perform atom counting using a resonant 461~nm fluorescence beam and a CCD camera.  We calibrate the fluorescence signal by measuring a vacuum Rabi splitting on the $^1$S$_0$ to $^3$P$_1$, $F = 9/2$ transition, as described in \cite{norcia2015strong}.  To measure total atom number, we apply pump lasers to the $^3$P$_0$ and $^3$P$_2$ to $^3$S$_1$ transitions to drive all atoms to the ground state. We then infer the number of atoms at the beginning of the 300~ms superradiance time window from a measured background loss rate from the lattice of $\gamma_0= 2.0(4)$/s and a measured additional atom-number dependent collissional loss rate of $ \gamma_N= 10(5) \times 10^{-6}$/(atom s) for excited state atoms. To account for the state-dependence of the collisions, we make a lowest order correction for the decrease in excited state population after the superradiant pulse, which leads to a 5\% correction at 175,000 atoms, and a \textless 1\% correction below 100,000 atoms.  


\subsection*{Simulations of Superradiance}
To provide theoretical predictions for superradiance in the presence of inhomogeneous coupling to the cavity mode and atomic decoherence and loss, we integrate a set of simplified optical Bloch equations:
$$\ \ \dot{J_z} = - C \gamma J_{\perp}^2 - \gamma_\ell J_z,\ \ \ \ \ \ \textrm{(S1)}$$
$$\dot{J_{\perp}} = (C\gamma J_z -\gamma_{\perp} -\gamma_\ell)J_{\perp}.$$

Here, $J_z$ and $J_{\perp}$ are defined in Figure 1, $\gamma_\ell$ accounts for atom loss, and $\gamma_{\perp}$ accounts for other homogeneous atomic decoherence.  In reality, both $\gamma_\ell$ and $\gamma_{\perp}$ depend on $N$, and on the instantaneous distribution of population in ground and excited states.  For simplicity, and to arrive at the analytic expressions given below, we assume $\gamma_\ell$ and $\gamma_{\perp}$ to be constants.  
In arriving at these equations, we have assumed that the cavity mode is on resonance with the atomic transition, and that the cavity mode occupation quickly equilibrates to the instantaneous value of $J_{\perp}$.  As discussed in \cite{Gross1982}, the dynamics of superradiance consist of an initial time period, during which quantum fluctuations are important in providing initial coherence, and a later time period during which the classically radiated field greatly exceeds quantum fluctuations and the dynamics proceed classically, governed by the above equations.  In the later time period, these equations can easily be solved for the case of homogeneous coupling and $\gamma_{\perp}, \gamma_\ell = 0$, with solutions of the form:
$$J_\perp(t) = \frac{N}{2}\mathrm{sech}\left(\frac{N C \gamma (t-t_d)}{2}\right), $$
$$J_z(t) = -\frac{N}{2}\tanh\left(\frac{N C \gamma (t-t_d)}{2}\right).$$

To simulate the quantum fluctuations in the cavity field that lead to the onset of superradiance, we include a drive to $J_{\perp}$ that represents random vacuum fluctuations that are replaced at the cavity amplitude decay rate $\kappa/2$.  This allows us to simulate the initial conditions that lead to the time delay $t_d$.  We verify that in the limit of large atom number our simulation reproduces the expected peak power, as well as the pulse duration $t_w$ and time delay $t_d$ of the peak output power derived in \cite{Gross1982}.

\subsection*{Effects of Atomic Decoherence and Loss}
Atomic decoherence, or a decay of $J_{\perp}$ at a rate $\gamma_{\perp}$, has the effect of setting a threshold atom number $N_t$ below which a superradiant pulse will not occur without seeding of coherence.  This threshold can be derived from Equation S1 by setting $J_z = N/2$ to determine the minimum value of $N$ for which fluctuations in $J_{\perp}$ will grow.  For atoms homogeneously coupled with cooperativity $C$, $N_t = 2 \gamma_\perp/(C \gamma)$.  For atom numbers above this threshold value, $N$ atoms radiating in the presence of decoherence produce a pulse identical to the pulse that the number of excess atoms, $N_x = N - N_t$, would produce in the absence of decoherence.  The solutions for $J_{\perp}$ and $J_z$ for homogeneous coupling in the presence of decoherence $\gamma_\perp$ are

$$ J_\perp(t) = \frac{N_x}{2}\mathrm{sech}\left(\frac{N_x C \gamma (t-t_d)}{2}\right), $$
$$ J_z(t) = -\frac{N_x}{2}\tanh\left(\frac{N_x C \gamma (t-t_d)}{2}\right) + N_t/2. $$

To treat the effects of atom loss, we rely on numerical simulation.  In the presence of atom loss at a rate $\gamma_\ell$, we observe a threshold-like behavior at an atom number of $N_{t\ell} \approx 13.6\gamma_\ell/(C \gamma)$, assuming homogeneous coupling.  For atom numbers well above threshold, peak output power and pulse duration are well-described by assuming an ensemble of $N-N_{t\ell}$ atoms radiating in the absence of loss.  Near threshold, however, the peak power radiated is higher and the pulse duration is shorter than that predicted by this analogy.  The delay time of the pulse exhibits a more striking difference: at fixed atom number, the delay time of the pulse peak increases linearly with loss rate up to the threshold loss rate.  

\def\arraystretch{2.5}
\begin{table*}[t]
  \centering
  \begin{tabular}{| l || l | l | l | l |}
    \hline
     & Homogeneous Coupling & Uniform Inhomogeneous Coupling \\ \hline \hline
    peak output rate $R_{peak}$                  & $\frac{1}{4}(N-N_t)^2 C \gamma$                         & $\approx \frac{1}{11.8}(N-N_t)^2 C \gamma $\\ \hline
    delay time $t_d$                 & $\approx \frac{(\ln{N} + \gamma_e)}{(N-N_t) C \gamma}$         & $\approx \frac{2(\ln{N} + \gamma_e)}{(N-N_t) C \gamma}$ \\ \hline
    duration $t_w$                & $\approx 3.5/(N C \gamma)$                                      & $\approx 7.05/(N C \gamma)$\\ \hline
    threshold atom number $N_t$             & $2 \gamma_\perp/C \gamma$         & $4 \gamma_\perp/C \gamma$ \\ \hline
    participation $M_{tot}/N$       & $(N-N_t)/N$         & $\approx 0.7(N-N_t)/N$ \\ 
    \hline
  \end{tabular}
  \label{S1}
\end{table*}

From our measured atom loss rate, assuming uniform inhomogeneous coupling (defined below), we expect a contribution to the threshold atom number of $3.0(6) \times 10^4$. The measured value is $3.3(8) \times 10^4$, which may indicate that the threshold behavior is due primarily to atom loss.  

The rates of atom loss and decoherence due to atomic collisions scale in proportion to the number of atoms, $N$.  This leads to a threshold atom number whose value depends on $N$.  This raises the threshold atom number compared to its value in the absence of $N$-dependent decoherence and leads to a constant fractional reduction in $N_x$. For our measured atom-number dependent loss rate, we expect the threshold atom number to increase by 0.13(7) for each atom added.  This effectively decreases $N_x$ by 13\% compared to its value if the same threshold were the result of atom number independent processes only.  Collisions that do not lead to atom loss, but whose rate would also scale with $N$ could lead to additional decrease in $N_x$.  

\subsection*{Effects of Inhomogeneous Coupling}
In our system, the atoms are trapped at anti-nodes of the 813~nm lattice, which are not aligned with the anti-nodes of the lasing cavity mode at 698~nm.  The coupling of an individual atom to the lasing mode thus depends on its location along the cavity axis.                   

We numerically simulate the effects of this inhomogeneous coupling by dividing the atoms into many classes, each with its own location $x_i$ and value of coupling to the lasing mode $g_i = g \cos(k x_i)$, where $k$ is the $k$-vector associated with the 698~nm cavity mode.  We may reasonably assume that before atoms are transferred to $^3$P$_0$, they are effectively uniformly distributed with respect to the phase of the standing-wave 698~nm cavity mode.  For simplicity, we analyze the case where the atoms maintain this uniform distribution after being transferred to $^3$P$_0$, meaning that the values of $x_i$ are sampled from a uniform distribution.  In reality, the adiabatic transfer is probably more effective for atoms located at the anti-nodes of the 698~nm cavity mode, as these experience a higher Rabi frequency from the transfer beam.  The true distribution of couplings for different transfer parameters will thus lie somewhere between the uniform distribution and the homogeneously coupled case.  

Below, we summarize the results of the numerical simulations for key parameters under these two conditions:  homogeneous coupling and uniformly inhomogeneous coupling.  Homogeneous coupling refers to the case where all $N$ atoms are at anti-nodes of the lasing mode and are coupled with peak coupling $g$.  Uniform inhomogeneous coupling refers to the case where $N$ atoms are distributed uniformly along the cavity mode.  In all expressions, $C$ stands for the peak cooperativity $C = \frac{4 g^2}{\kappa \gamma}$.

The effect of uniform inhomogeneous coupling on $t_d$, $t_w$, and $N_t$ is to reduce $C$ by a factor of 2, corresponding to taking the spatial average of $C$ over the ensemble of atoms.  Its effects on the peak photon output rate $R_{peak}$ and the fraction of atoms emitted into the cavity mode $M_{tot}/N$, however, are more complex.  Differing Rabi frequencies within the ensemble cause the atoms to dephase with respect to the polar angle $\theta$.  This shortens the length of the collective Bloch vector at the time of peak emission and strands poorly coupled atoms in $\ket{e}$ when the pulse terminates, leaving 30\% of atoms in the excited state.  

The finite temperature of the atoms within the lattice leads to a reduction of $C$ through several mechanisms.  The finite temperature of the atoms leads to a non-zero radial extent of the ensemble, which causes the atoms to sample regions with lower coupling to the cavity mode.  At 10~$\mu$K, this leads to a 10\% reduction in the spatially averaged $C$.  The finite Lamb-Dicke parameter $\eta = 0.16$ also leads to a reduction $C$ for the motional carrier transition observed here.  We estimate this effect to lead to a 7.5\% reduction in $C$.  To compare measured values of $R_{peak}$ and $t_w$ to predictions in the main text, we take into account these two temperature-related effects, as well as the 13\% reduction in $N_x$ from the measured atom number dependent loss rate.

\end{document}